# Effects of transverse electron dispersion on photo-emission spectra of quasi-one-dimensional systems

Željana Agić[a], Paško Županović[a], Aleksa Bjeliš[b]

[a]Faculty of Natural Sciences, Mathematics and Education, University of Split, N. Tesle 12, 21 000 Split, Croatia; e-mail: agicz@pmfst.hr; pasko@pmfst.hr
[b]Faculty of Science, University of Zagreb, Bijenička 32, 10000 Zagreb, Croatia; e-mail: bjelis@phy.hr

**Abstract.** The random phase approximation (RPA) spectral function of the one-dimensional electron band with the three dimensional long range Coulomb interaction shows a broad feature which is spread on the scale of the plasmon energy and vanishes at the chemical potential. The fact that there are no quasi-particle δ-peaks is the direct consequence of the acoustic nature of the collective plasmon mode. This behaviour of the spectral function is in the qualitative agreement with the angle resolved photo-emission spectra of some Bechgaard salts. In the present work we consider the modifications in the spectral function due to finite transverse electron dispersion. The transverse bandwidth is responsible for the appearance of an optical gap in the long wavelength plasmon mode. The plasmon dispersion of such kind introduces the quasi-particle δ-peak into the spectral function at the chemical potential. The cross-over from the Fermi liquid to the non-Fermi liquid regime by decreasing the transverse bandwidth takes place through the decrease of the quasi-particle weight as the optical gap in the long wavelength plasmon mode is closing.

The angle resolved photoemission (ARPES) experiments in some quasi-one-dimensional materials, particularly in Bechgaard salts like $(TMTSF)_2 PF_6$, show a wide feature in the spectral density at frequency scale of the order of plasmon frequency, and do not show quasi-particle peaks, main ingredients in spectra of three-dimensional Fermi liquids [1]. Similar features are also observed in some quasi-two-dimensional high-$T_c$ superconductors whose ARPES spectra have the so-called peak-dip-hump structure where the hump spreads up to energies far from the Fermi surface [2]. As far as wide frequency scale is considered the measurements [1] are in qualitative agreement with our RPA result for strictly one-dimensional band dispersion [3]. Physically the origin of wide feature in the spectral density $A(\mathbf{k},\omega)$ is a plasmon collective mode which, due to its anisotropic acoustic dispersion, prevents the appearance of quasi-particle peaks. Note that the range of small frequencies (close to the Fermi energy) belongs to the Luttinger liquid regime and is beyond the RPA treatment.

The aim of the present work is to investigate the effects of a finite transverse band dispersion (roughly measured by the hopping integral between neighbouring chains, $t$) on the spectral density. Generally, as $t$ increases one expects a cross-over to a three-dimensional regime, and finally an asymptotic tendency towards the spectral density for the isotropic band dispersion calculated in early work of Hedin and Lundqvist [4]. To determine detail of this cross-over we extend our previous calculation of the reciprocal electron propagator in RPA

$$G^{-1}(\mathbf{k},\omega) = G_0^{-1}(\mathbf{k},\omega) - \frac{i}{2\pi N}\sum_{\mathbf{q}}\left[\int d\omega' \overline{V}(\mathbf{q},\omega')G_0(\mathbf{k}-\mathbf{q},\omega-\omega') + i\pi V(\mathbf{q})\right] \qquad (1)$$

by including anisotropic band dispersion $E(\mathbf{k}) = -2t_0(\cos k_\| b - \cos k_F b) - 2t(\cos k_x a + \cos k_z c)$. With finite $t$ we have a finite transverse plasmon frequency $\omega_p^2 = 16e^2 t^2 / v_F$ entering into plasmon dispersion $\omega^2(\mathbf{q}) = (\Omega_p^2 q_\|^2 + \omega_p^2 q_\perp^2)/q^2$, dielectric function $\varepsilon_m(\mathbf{q},\omega) = 1 - \omega^2(\mathbf{q})/\omega^2$ and screened Coulomb potential $\overline{V}(\mathbf{q},\omega) = V(\mathbf{q})/\varepsilon_m(\mathbf{q},\omega)$. We simplify further steps in integrating Eq.1 by taking the limit $t<<\omega_p$ and neglecting the longitudinal band dispersion $v_F q_\parallel$ with respect to $\Omega_p$. These approximations cause a loss of spectral density for small $\Omega_p$, but do not change its main qualitative properties.

The resulting spectral function $A(\mathbf{k},\omega)$ for two values of transverse plasmon frequency $\omega_p$ is shown in Fig.1. The quasi-particle peaks are present for any finite $t$, and are superposed to a slightly modified wide feature present already for $t=0$. The appearance of quasi-particle $\delta$-peaks is directly related to an optical gap in the long wavelength plasmon dispersion opened due to a finite value of $t$. Then the interval $-\omega_p + E_0(k_\parallel) < \omega < \omega_p + E_0(k_\parallel)$ always contains a value $\omega_0(\mathbf{k})$ for which $\mathrm{Re}\,G^{-1}[\mathbf{k},\omega_0(\mathbf{k})]=$ $\mathrm{Im}\,G^{-1}[\mathbf{k},\omega_0(\mathbf{k})]=0$. Then the spectral function can be written as $A(\mathbf{k},\omega) = Z(\mathbf{k})\delta(\omega - \omega_0(\mathbf{k}))$ where $Z(\mathbf{k}) = \left|\partial \mathrm{Re}\,G^{-1}(\mathbf{k},\omega_0(\mathbf{k}))/\partial\omega\right|^{-1}$ is the weight of the quasi-particle. The dependence of $Z(\mathbf{k})$ on $t$ for $\mathbf{k}$ at the Fermi surface is shown in Fig 2. The weight of the quasi particle peak increases as $t$ increases.

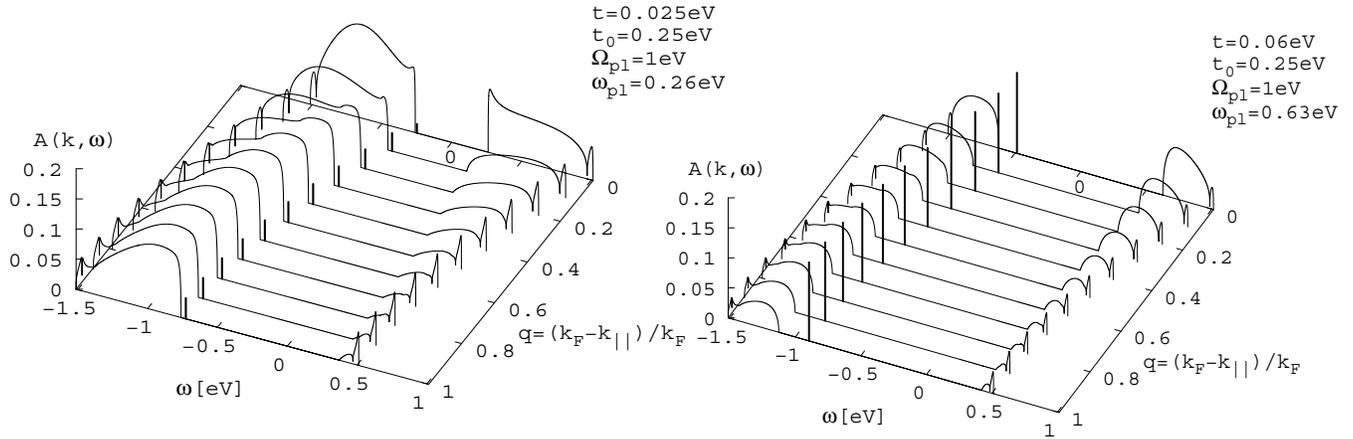

(a)　　　　　　　　　　　　　　　　　　(b)

**Figure 1.** Spectral function for small (a) and large (b) value of transverse plasmon frequency $\omega_p$.

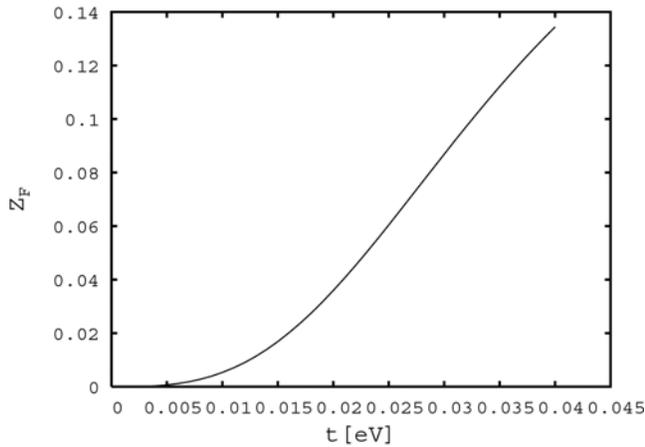

**Figure 2.** The quasi-particle weight Z on the Fermi surface.

The above results suggest that in $(\mathrm{TMTSF})_2\mathrm{PF}_6$ (for which $t\approx 0.0125\,\mathrm{eV}$) the quasi particle peak has a weight of the order of 1%, and is positioned about 7meV from the wide feature in the spectral density. This is of the order of the resolution of the ARPES experiments, and is therefore hardly visible.

We conclude that by increasing the transverse bandwidth the cross-over from 1d to 3d regime in the spectral density takes place through a gradual increase of the quasi-particle peak weight on account of the weight of a broad feature caused by the plasmon excitations. The peak simultaneously moves away from the broad feature on the energy scale.

**Acknowledgements**
The present work is supported by Croatian Ministry of Science and Technology, project number 0119251.